\documentclass[twocolumn,showpacs,preprintnumbers,showkeys,amsmath,amssymb]{revtex4}
\usepackage{graphicx}
\usepackage{amssymb}
\usepackage{amsmath}
\usepackage{bm}
\usepackage[flushleft]{caption2}

\begin{document}
\setcounter{page}{1}
\title{Influence of velocity curl on conservation laws}
\author{Zihua Weng}
\email{xmuwzh@xmu.edu.cn.} \affiliation{School of Physics and
Mechanical \& Electrical Engineering,
\\Xiamen University, Xiamen 361005, China}

\begin{abstract}
The paper discusses impact of the velocity curl on some conservation
laws in the gravitational field and electromagnetic field, by means
of the characteristics of quaternions. When the velocity curl can
not be neglected, it will cause the predictions to departure
slightly from the conservation laws, which include mass continuity
equation, charge continuity equation, and conservation of spin, etc.
And the scalar potential of gravitational field has an effect on the
speed of light, the conservation of mass, and conservation of
charge, etc. The results explain how the velocity curl influences
some conservation laws in the gravitational field and
electromagnetic field.
\end{abstract}



\maketitle

\section{INTRODUCTION}

In the gravitational field, the rotating physical object can be
considered as the particle with the velocity curl. But the existing
theories can not deal with the theoretical model. The paper attempts
to solve this problem, and reason out why the angular velocity has
an influence on the conservation laws about the particle motion.

The quaternion \cite{hamilton} can be used to describe the property
of the gravitational field, including the scalar invariants and
conservation laws. In the gravitational field, there exist the
conservation of mass \cite{lavoisier}, the mass invariant, the
energy invariant \cite{young}, and the conservation of energy.
Making use of the scalar property of the quaternion, the potential
and the strength of gravitational field both have the influence on
these conservations. While the angular velocity has an impact on
some physical quantities, such as the gyroscopic torque and Coriolis
force.

With the features of the quaternions, we find the mass density, the
spin density \cite{uhlenbeck}, and the energy density may be
variable respectively under the quaternion coordinate transformation
in the gravitational field \cite{weng} and electromagnetic field. So
are the mass continuity equation, the charge continuity equation,
the spin continuity equation, and the energy continuity equation,
and the continuity equation of spin magnetic moment, etc.

\section{Scalar invariants of gravitational field}

Making use of the characteristics of the quaternions, we may obtain
some kinds of scalar invariants in the case for coexistence of
gravitational field and electromagnetic field with the velocity
curl, under the octonion coordinate transformation. And each
octonion definition of physical quantity possesses one invariant
equation.

\subsection{Quaternion transformation}

In the quaternion space, the coordinates $d_0,~d_1,~d_2$, and $d_3$
are all real, the basis vector is $\mathbb{E}_g = (1,
\emph{\textbf{i}}_1, \emph{\textbf{i}}_2, \emph{\textbf{i}}_3)$. The
quaternion $\mathbb{D}(d_0, d_1, d_2, d_3)$ is defined as,
\begin{equation}
\mathbb{D} = d_0 + d_1 \emph{\textbf{i}}_1 + d_2 \emph{\textbf{i}}_2
+ d_3 \emph{\textbf{i}}_3
\end{equation}

When the coordinate system is transformed into the other, the
quaternion $\mathbb{D}$ will be become to the physical quantity
$\mathbb{D}' (d'_0 , d'_1 , d'_2 , d'_3)$.
\begin{equation}
\mathbb{D}' = \mathbb{K}^* \circ \mathbb{D} \circ \mathbb{K}
\end{equation}
where, the $\mathbb{K}$ is the quaternion, and $\mathbb{K}^* \circ
\mathbb{K} = 1$; the $\circ$ denotes the quaternion multiplication;
the $*$ indicates the conjugate of quaternion.

Both sides' scalar parts keep unchanged in the above, when the
quaternion coordinate system is transforming. Therefore
\begin{eqnarray}
d_0 = d'_0
\end{eqnarray}

Making use of the above equation, we can obtain some kinds of scalar
invariant equations under the quaternion coordinate transformation
in the gravitational field.

\subsection{Radius vector}

In the quaternion space, the coordinates are $r_0 , r_1 , r_2$, and
$r_3$, while the radius vector is $\mathbb{R}_g = \Sigma(r_i
\emph{\textbf{i}}_i)$. It can be combined with the quaternion
$\mathbb{X}_g = \Sigma(x_i \emph{\textbf{i}}_i)$ to become the
compounding radius vector
\begin{eqnarray}
\mathbb{\bar{R}}_g = \mathbb{R}_g + k_{rx} \mathbb{X}_g~.
\end{eqnarray}
where, $r_0 = v_0 t$; $x_0 = a_0 t$; $v_0$ is the speed of light
beam; $a_0$ is the scalar potential of gravitational field; $t$
denotes the time; $k_{rx} = 1$ and $k_{eg}$ are the coefficients.

In other words, the $\mathbb{\bar{R}}_g$ can be considered as the
radius vector in the quaternion compounding space, with the basis
vector $(\emph{\textbf{i}}_0, \emph{\textbf{i}}_1,
\emph{\textbf{i}}_2, \emph{\textbf{i}}_3)$, and $\emph{\textbf{i}}_0
= 1$.

When the coordinate system is transformed into the other, we have a
radius vector $\mathbb{\bar{R}}_g'(\bar{r}'_0, \bar{r}'_1,
\bar{r}'_2, \bar{r}'_3)$, with the $\bar{r}'_0 = (v'_0 + k_{rx}
a'_0) t'$ from Eq.(2).

From Eqs.(3) and (4), we obtain
\begin{eqnarray}
\bar{r}_0 = \bar{r}'_0
\end{eqnarray}
where, $\bar{r}_0 = \bar{v}_0 t$ ; $\bar{v}_0 = v_0 + k_{rx} a_0$ .

In some special cases, we may substitute a quaternion quantity
$\mathbb{\bar{Z}}_g(\bar{z}_0, \bar{z}_1, \bar{z}_2, \bar{z}_3)$ for
the quaternion radius vector $\mathbb{\bar{R}}_g(\bar{r}_0,
\bar{r}_1, \bar{r}_2, \bar{r}_3)$. The former quantity is defined as
\begin{equation}
\mathbb{\bar{Z}}_g = \mathbb{\bar{R}}_g \circ \mathbb{\bar{R}}_g
\end{equation}
where, $\bar{z}_0 = (\bar{r}_0)^2 - (\bar{r}_1)^2 - (\bar{r}_2)^2 -
(\bar{r}_3)^2$ .

When the quaternion coordinate system is rotated, the physical
quantity $\mathbb{\bar{Z}}_g$ is become into the physical quantity
$\mathbb{\bar{Z}}_g'(\bar{z}'_0, \bar{z}'_1, \bar{z}'_2,
\bar{z}'_3)$ from Eq.(2).

From Eqs.(3) and (6), we gain
\begin{eqnarray}
(\bar{r}_0)^2 - \Sigma (\bar{r}_j)^2 = (\bar{r}'_0)^2 - \Sigma
(\bar{r}'_j)^2
\end{eqnarray}
where, $j = 1, 2, 3$; $i = 0, 1, 2, 3 $.

The above means that there may exist the special case of the $x_i
\neq 0$ when the $r_i = 0$. We can obtain different invariants from
the physical definitions, by means of the characteristics of
quaternion coordinate transformation in the gravitational field.

\begin{table}[t]
\caption{\label{tab:table1}The quaternion multiplication table.}
\begin{ruledtabular}
\begin{tabular}{ccccc}
$ $ & $1$ & $\emph{\textbf{i}}_1$  & $\emph{\textbf{i}}_2$ &
$\emph{\textbf{i}}_3$  \\
\hline $1$ & $1$ & $\emph{\textbf{i}}_1$  & $\emph{\textbf{i}}_2$ &
$\emph{\textbf{i}}_3$  \\
$\emph{\textbf{i}}_1$ & $\emph{\textbf{i}}_1$ & $-1$ &
$\emph{\textbf{i}}_3$  & $-\emph{\textbf{i}}_2$ \\
$\emph{\textbf{i}}_2$ & $\emph{\textbf{i}}_2$ &
$-\emph{\textbf{i}}_3$ & $-1$ & $\emph{\textbf{i}}_1$ \\
$\emph{\textbf{i}}_3$ & $\emph{\textbf{i}}_3$ &
$\emph{\textbf{i}}_2$ & $-\emph{\textbf{i}}_1$ & $-1$
\end{tabular}
\end{ruledtabular}
\end{table}

\subsection{Speed of light}

In the quaternion space, the velocity $\mathbb{V}_g (v_0, v_1, v_2,
v_3)$ is combined with gravitational potential $\mathbb{A}_g (a_0,
a_1, a_2, a_3)$ to become the compounding velocity
$\mathbb{\bar{V}}_g (\bar{v}_0, \bar{v}_1, \bar{v}_2, \bar{v}_3)$.
\begin{eqnarray}
\mathbb{\bar{V}}_g = \mathbb{V}_g + k_{rx} \mathbb{A}_g
\end{eqnarray}
where, the component $a_j = 0$ in Newtonian gravity.

When the quaternion coordinate system is transformed into the other,
we have one velocity $\mathbb{\bar{V}}_g' (\bar{v}'_0, \bar{v}'_1,
\bar{v}'_2, \bar{v}'_3 )$ from Eq.(2) in the quaternion compounding
space.

From Eq.(3) and the definition of velocity, we obtain the invariant
equation about the speed of light.
\begin{eqnarray}
\bar{v}_0 = \bar{v}'_0
\end{eqnarray}

Choosing the definition combination of the quaternion velocity and
radius vector, we find the invariants, Eqs.(5) and (9), and the
Galilean transformation.
\begin{eqnarray}
\bar{r}_0 = \bar{r}'_0~,~\bar{v}_0 = \bar{v}'_0~.
\end{eqnarray}

In some cases, we obtain Lorentz transformation from the invariant
equations, Eqs.(7) and (9).
\begin{eqnarray}
(\bar{r}_0)^2 - \Sigma (\bar{r}_j)^2 = (\bar{r}'_0)^2 - \Sigma
(\bar{r}'_j)^2~,~\bar{v}_0 = \bar{v}'_0~.
\end{eqnarray}

The above states the speed of light, $v_0$, may be variable in some
cases, due to the influence of scalar potential of gravitational
field \cite{hau}. And we can obtain some kinds of coordinate
transformations from the different definition combinations in
quaternion compounding space, such as the Galilean and Lorentz
transformations, etc.

\subsection{Potential and strength}

In the quaternion compounding space, the potential
$\mathbb{\bar{A}}_g (\bar{a}_0, \bar{a}_1, \bar{a}_2, \bar{a}_3)$ is
defined from Eq.(8).
\begin{eqnarray}
\mathbb{\bar{A}}_g = \mathbb{A}_g + K_{rx} \mathbb{V}_g
\end{eqnarray}
where, $K_{rx} = 1 / k_{rx}$; $\mathbb{A}_g$ is the gravitational
potential.

The strength $\mathbb{\bar{B}}_g (\bar{b}_0, \bar{b}_1, \bar{b}_2,
\bar{b}_3)$ is defined from Eq.(12).
\begin{eqnarray}
\mathbb{\bar{B}}_g = \lozenge \circ \mathbb{\bar{A}}_g =
\mathbb{B}_g + K_{rx} \mathbb{U}_g
\end{eqnarray}
where, $\lozenge = \Sigma (\emph{\textbf{i}}_i
\partial_i)$, with $\partial_i = \partial/\partial \bar{r}_i$; $\nabla = \Sigma
(\emph{\textbf{i}}_j \partial_j)$; in most cases, $\bar{r}_i \approx
r_i$, and then $\partial/\partial \bar{r}_i \approx
\partial/\partial r_i$; the velocity $\mathbb{V}_g = v_0 [\lozenge \circ \mathbb{R}_g - \nabla \cdot
\left\{ \Sigma (r_j \emph{\textbf{i}}_j ) \right\} ]$, velocity curl
$\mathbb{U}_g = \lozenge \circ \mathbb{V}_g$; $\mathbb{A}_g = v_0
\lozenge \circ \mathbb{X}_g$; $\mathbb{B}_g = \lozenge \circ
\mathbb{A}_g$; $\bar{b}_0 = \partial_0 \bar{a}_0 - \Sigma
(\partial_j \bar{a}_j)$.

In the planar polar coordinates, the velocity $\textbf{v} =
\vec{\omega} \times \textbf{r}$, with $\vec{\omega}$ being the
angular velocity. And then, we have the relation $\nabla \times
\textbf{v} = 2 \vec{\omega}$ . Where, $\textbf{v} = \Sigma (v_j
\emph{\textbf{i}}_j)$.

In the paper, we choose the gauge condition, $\bar{b}_0 = 0$, to
simplify succeeding calculation. And the gravitational strength is
written as $\mathbb{\bar{B}}_g = \overline{\textbf{g}}/\bar{v}_0 +
\overline{\textbf{b}}$ .
\begin{eqnarray}
\overline{\textbf{g}}/\bar{v}_0 = && \partial_0
\overline{\textbf{a}} + \nabla \bar{a}_0
\\
\overline{\textbf{b}} = && \nabla \times \overline{\textbf{a}}
\end{eqnarray}
where, $\overline{\textbf{a}} = \Sigma (\bar{a}_j
\emph{\textbf{i}}_j)$; $\overline{\textbf{b}} =  \Sigma (\bar{b}_j
\emph{\textbf{i}}_j)$.

When the quaternion coordinate system is transformed into the other,
we have the potential $\mathbb{\bar{A}}_g' (\bar{a}'_0, \bar{a}'_1,
\bar{a}'_2, \bar{a}'_3 )$ and strength $\mathbb{\bar{B}}_g'
(\bar{b}'_0, \bar{b}'_1, \bar{b}'_2, \bar{b}'_3 )$ respectively from
Eq.(2).

From Eqs.(3) and (12), we will obtain the invariant equation about
the scalar potential,
\begin{eqnarray}
\bar{a}_0 = \bar{a}'_0
\end{eqnarray}
and the strength invariant from Eqs.(3) and (13).
\begin{eqnarray}
\bar{b}_0 = \bar{b}'_0
\end{eqnarray}

The above equations state the gravitational field in the quaternion
compounding space possesses not only the potential and strength but
also the velocity and velocity curl. While, the scalar potential
$a_0$ and scalar strength $b_0$ of the gravitational field are
variable, due to the influence of the velocity and the velocity
curl.

\subsection{Conservation of mass}

The source density $\mathbb{\bar{S}}'$ is defined from the strength
$\mathbb{\bar{B}}_g$ in the quaternion compounding space.
\begin{eqnarray}
\mu \mathbb{\bar{S}}' && = - ( \mathbb{\bar{B}}_g/\bar{v}_0 +
\lozenge)^* \circ \mathbb{\bar{B}}_g
\nonumber\\
&& = \mu_g^g \mathbb{\bar{S}}_g - \mathbb{\bar{B}}_g^* \circ
\mathbb{\bar{B}}_g/\bar{v}_0
\end{eqnarray}
where, $\mu$ and $\mu_g^g$ are the constants; $\mathbb{\bar{S}}'$
covers the linear momentum density $\mathbb{\bar{S}}_g$ and an extra
part $ \mathbb{\bar{B}}_g^* \circ \mathbb{\bar{B}}_g/(\bar{v}_0
\mu_g^g) $; and $\mathbb{\bar{B}}_g^* \circ
\mathbb{\bar{B}}_g/(2\mu_g^g)$ is the energy density of
gravitational field, and is similar to that of electromagnetic
field.

The linear momentum density $\mathbb{\bar{P}}_g(\bar{p}_0,
\bar{p}_1, \bar{p}_2, \bar{p}_3)$ is the extension of
$\mathbb{\bar{S}}_g = m \mathbb{\bar{V}}_g$.
\begin{eqnarray}
\mathbb{\bar{P}}_g = \mu \mathbb{\bar{S}}' / \mu_g^g
\end{eqnarray}
where, $\bar{p}_0 = \widehat{m} \bar{v}_0$ , $\bar{p}_j = m
\bar{v}_j \emph{\textbf{i}}_j$ ; $\widehat{m} = m + \triangle m $;
$m$ is the inertial mass density, and $\widehat{m}$ is the
gravitational mass density; $\triangle m = - \mathbb{\bar{B}}_g^*
\circ \mathbb{\bar{B}}_g/(\bar{v}_0^2 \mu_g^g)$.

When the quaternion coordinate system rotates, we have the linear
momentum density $\mathbb{\bar{P}}_g'(\bar{p}'_0, \bar{p}'_1,
\bar{p}'_2, \bar{p}'_3)$, and the scalar invariant equation from
Eqs.(3) and (19).
\begin{eqnarray}
\widehat{m} \bar{v}_0 = \widehat{m}' \bar{v}'_0
\end{eqnarray}

By Eqs.(9) and (20), the gravitational mass density $\widehat{m}$
remains unchanged in the gravitational field. And then we have the
conservation of mass as follows.
\begin{eqnarray}
\widehat{m} = \widehat{m}'
\end{eqnarray}

In the same way, the inertial mass density $m$ is the invariant from
the definitions of $\mathbb{\bar{V}}_g$ and $\mathbb{\bar{S}}_g$ .
\begin{eqnarray}
m = m'
\nonumber
\end{eqnarray}

The above can also be reduced from Eq.(21), in the case of the
$\bar{b}_j = 0$ and $\triangle m = 0$.

The above means that if we emphasize the definitions of the velocity
and linear momentum, the gravitational mass density will remain the
same under the coordinate transformation in Eq.(2) in the
compounding space. The results are the same as that in the
quaternion space.

\subsection{Mass continuity equation}

The applied force density $\mathbb{\bar{F}}_g(\bar{f}_0, \bar{f}_1,
\bar{f}_2, \bar{f}_3)$ is defined from the linear momentum density
$\mathbb{\bar{P}}_g$ .
\begin{eqnarray}
\mathbb{\bar{F}}_g = \bar{v}_0 (\mathbb{\bar{B}}_g / \bar{v}_0 +
\lozenge )^* \circ \mathbb{\bar{P}}_g
\end{eqnarray}
where, the applied force density includes the inertial force density
and the gravitational force density, etc. While, the scalar
$\bar{f}_0 = \bar{v}_0 \partial \bar{p}_0 / \partial r_0 + \bar{v}_0
\Sigma (\partial \bar{p}_j / \partial r_j ) + \Sigma ( \bar{b}_j
\bar{p}_j )$.

From the above, the cross product of velocity curl with linear
momentum is the part of force, which is similar to the Coriolis
force \cite{coriolis} in Newtonian gravity.

When the quaternion coordinate system rotates, we have the force
density $\mathbb{\bar{F}}_g' (\bar{f}'_0, \bar{f}'_1, \bar{f}'_2,
\bar{f}'_3 )$ from Eqs.(2), (3) and (22). And then, we have
\begin{eqnarray}
\bar{f}_0 = \bar{f}'_0
\end{eqnarray}

In the equilibrium state, there is $\mathbb{\bar{F}}_g = 0$. And we
have the mass continuity equation by Eqs.(9) and (23).
\begin{eqnarray}
\partial \bar{p}_0 / \partial r_0 + \Sigma (  \partial \bar{p}_j /
\partial r_j ) + \Sigma ( \bar{b}_j \bar{p}_j ) / \bar{v}_0 = 0
\end{eqnarray}
further, if the $a_j = \bar{b}_j = 0$, the above is reduced to
\begin{eqnarray}
\partial m / \partial t + \Sigma ( \partial p_j /
\partial r_j ) = 0
\end{eqnarray}

The above states the potential $\mathbb{\bar{A}}_g$, strength
$\mathbb{\bar{B}}_g$, and curl $\mathbb{U}_g$ have influences on the
mass continuity equation, although the term $\Sigma ( \bar{b}_j
\bar{p}_j ) / \bar{v}_0$ and extra mass density $\triangle m$ both
are very tiny. The mass continuity equation is the invariant under
the quaternion transformation, if we choose the definition
combination of the velocity and applied force in the gravitational
field.

\begin{table}[b]
\caption{\label{tab:table1}The physical quantities of gravitational
field in the quaternion compounding space.}
\begin{ruledtabular}
\begin{tabular}{cccc}
$space$               &    $field$               &    $mechanics$          &    $matter$               \\
\hline
$\mathbb{\bar{R}}_g$  &    $\mathbb{\bar{X}}_g$  &    $\mathbb{\bar{P}}_g$ &    $\mathbb{\bar{L}}_g$   \\
$\mathbb{\bar{V}}_g$  &    $\mathbb{\bar{A}}_g$  &    $\mathbb{\bar{F}}_g$ &    $\mathbb{\bar{W}}_g$   \\
$\mathbb{\bar{U}}_g$  &    $\mathbb{\bar{B}}_g$  &    $\mathbb{\bar{C}}_g$ &    $\mathbb{\bar{N}}_g$   \\
\end{tabular}
\end{ruledtabular}
\end{table}

\subsection{$\bar{f}_0$ continuity equation}

In a similar way, we obtain a new physical quantity density,
$\mathbb{\bar{C}}_g(\bar{c}_0, \bar{c}_1, \bar{c}_2, \bar{c}_3)$,
which can be defined from the applied force density
$\mathbb{\bar{F}}_g$ .
\begin{eqnarray}
\mathbb{\bar{C}}_g = \bar{v}_0 (\mathbb{\bar{B}}_g/\bar{v}_0 +
\lozenge )^* \circ \mathbb{\bar{F}}_g
\end{eqnarray}
where, $\bar{c}_0 = \bar{v}_0 \partial \bar{f}_0 /
\partial r_0 + \bar{v}_0 \Sigma (\partial \bar{f}_j / \partial r_j )
+ \Sigma ( \bar{b}_j \bar{f}_j ) $ is the scalar part of
$\mathbb{\bar{C}}_g$.

When the quaternion coordinate system rotates, we have the density
$\mathbb{\bar{C}}_g' (\bar{c}'_0, \bar{c}'_1, \bar{c}'_2, \bar{c}'_3
)$ from Eqs.(2), (3), and (26). And then, we have
\begin{eqnarray}
\bar{c}_0 = \bar{c}'_0 ~.
\end{eqnarray}

In some cases, there is $\mathbb{\bar{C}}_g = 0$. And then we have
the continuity equation about $\bar{f}_0$ by Eqs.(9) and (27).
\begin{eqnarray}
\partial \bar{f}_0 / \partial r_0 + \Sigma (  \partial \bar{f}_j /
\partial r_j ) + \Sigma ( \bar{b}_j \bar{f}_j ) / \bar{v}_0 = 0
\end{eqnarray}
further, if the $\bar{b}_j = 0$, the above is reduced to
\begin{eqnarray}
\partial \bar{f}_0 / \partial r_0 + \Sigma (  \partial \bar{f}_j /
\partial r_j ) = 0.
\end{eqnarray}

The above states the strength $\mathbb{\bar{B}}_g$ and velocity curl
$\mathbb{U}_g$ have an influence on the continuity equation about
$\bar{f}_0$, although we do not know much about the physical
quantity $\mathbb{\bar{C}}_g$. This continuity equation is also the
invariant under the quaternion transformation.

\subsection{Spin density}

The angular momentum density $\mathbb{\bar{L}}_g(\bar{l}_0,
\bar{l}_1, \bar{l}_2, \bar{l}_3)$ can be defined from the linear
momentum density $\mathbb{\bar{P}}_g$ and radius vector
$\mathbb{\bar{R}}_g$ in the quaternion compounding space.
\begin{eqnarray}
\mathbb{\bar{L}}_g = \mathbb{\bar{R}}_g \circ \mathbb{\bar{P}}_g
\end{eqnarray}
where, the scalar part $\bar{l}_0 = \bar{r}_0 \bar{p}_0 - \Sigma
(\bar{r}_j \bar{p}_j )$.

The scalar $\bar{l}_0$ is regarded as the density of spin angular
momentum in the quaternion compounding space. And the $\bar{l}_0$
covers the spin angular momentum density, $l_0$, in the
gravitational field.

When the quaternion coordinate system rotates, we have the angular
momentum density $\mathbb{\bar{L}}_g' (\bar{l}'_0 , \bar{l}'_1 ,
\bar{l}'_2 , \bar{l}'_3 )$ from Eq.(2). And then we have the
conservation of spin from the above and Eq.(3).
\begin{eqnarray}
\bar{l}_0 = \bar{l}'_0
\end{eqnarray}

The above means that the velocity and gravitational strength both
have the influence on the orbital angular momentum and the spin
angular momentum. And the spin angular momentum, $l_0$, will be
variable under the quaternion transformation in Eq.(2) in the
quaternion compounding space.

\begin{table}[b]
\caption{\label{tab:table1}The definitions of scalar invariants of
gravitational field in the quaternion compounding space.}
\begin{ruledtabular}
\begin{tabular}{lll}
$definition$            &    $invariant $                &    $ meaning $                            \\
\hline
$\mathbb{\bar{R}}_g$    &    $\bar{r}_0 = \bar{r}'_0 $   &    $Galilean~invariant$                   \\
$\mathbb{\mathbb{\bar{R}}}_g \circ \mathbb{\bar{R}}_g$   &    $\bar{z}_0 = \bar{z}'_0 $  &  $Lorentz~invariant$  \\
$\mathbb{\bar{V}}_g$    &    $\bar{v}_0 = \bar{v}'_0$    &    $invariable~speed~of~light$            \\
$\mathbb{\bar{A}}_g$    &    $\bar{a}_0 = \bar{a}'_0$    &    $invariable~scalar~potential$          \\
$\mathbb{\bar{B}}_g$    &    $\bar{b}_0 = \bar{b}'_0$    &    $invariable~gauge$                     \\
$\mathbb{\bar{P}}_g$    &    $\bar{p}_0 = \bar{p}'_0$    &    $conservation~of~mass$                 \\
$\mathbb{\bar{F}}_g$    &    $\bar{f}_0 = \bar{f}'_0$    &    $mass~continuity~equation$             \\
$\mathbb{\bar{L}}_g$    &    $\bar{l}_0 = \bar{l}'_0$    &    $invariable~spin~density$              \\
$\mathbb{\bar{W}}_g$    &    $\bar{w}_0 = \bar{w}'_0$    &    $conservation~of~energy$               \\
$\mathbb{\bar{N}}_g$    &    $\bar{n}_0 = \bar{n}'_0$    &    $energy~continuity~equation$           \\
\end{tabular}
\end{ruledtabular}
\end{table}

\subsection{Conservation of energy}

The total energy density $\mathbb{\bar{W}}_g(\bar{w}_0, \bar{w}_1,
\bar{w}_2, \bar{w}_3)$ is defined from the angular momentum density
$\mathbb{\bar{L}}_g$ in the quaternion compounding space.
\begin{eqnarray}
\mathbb{\bar{W}}_g  = \bar{v}_0 ( \mathbb{\bar{B}}_g/\bar{v}_0 +
\lozenge) \circ \mathbb{\bar{L}}_g
\end{eqnarray}
where, the scalar part $\bar{w}_0 = \overline{\textbf{b}} \cdot
\overline{\textbf{l}} + \bar{v}_0 \partial_0 \bar{l}_0 + \bar{v}_0
\nabla \cdot \overline{\textbf{l}}$ is the energy density;
$\overline{\textbf{w}} = \Sigma (\bar{w}_j \emph{\textbf{i}}_j )$; $
\overline{\textbf{l}} = \Sigma (\bar{l}_j \emph{\textbf{i}}_j)$.

The total energy incorporates the potential energy, the kinetic
energy, the torque, and the work, etc.

In the above, the cross product of angular momentum with velocity
curl is the part torque, which is similar to the gyroscopic torque
in Newtonian gravity. While the cross product of angular momentum
with gravitational strength is the part torque also.

When the quaternion coordinate system rotates, we have the total
energy density $\mathbb{\bar{W}}_g' (\bar{w}'_0, \bar{w}'_1,
\bar{w}'_2, \bar{w}'_3 )$ from Eq.(2). And then, we have the
conservation of energy by Eq.(3) and the above.
\begin{eqnarray}
\bar{w}_0 = \bar{w}'_0
\end{eqnarray}

In some special cases, there exists $\bar{w}'_0 = 0$, we obtain the
spin continuity equation by Eqs.(9) and Eq.(33).
\begin{eqnarray}
\partial \bar{l}_0 / \partial r_0 + \nabla \cdot \overline{\textbf{l}}
+ \overline{\textbf{b}} \cdot \overline{\textbf{l}} / \bar{v}_0 = 0
\end{eqnarray}

If the last term is neglected, the above is reduced to
\begin{eqnarray}
\partial \bar{l}_0 / \partial r_0 + \nabla \cdot \overline{\textbf{l}} =
0~.
\end{eqnarray}
where, when the time $t$ is only the independent variable, the
$\partial \bar{l}_0 / \partial t$ will become the $d \bar{l}_0 / d
t$ .

The above means that the energy density, $w_0$, may be variable
under the quaternion transformation in Eq.(2) in the gravitational
field, because of the gravitational strength and the velocity curl
have the influence on the angular momentum density.

\subsection{Energy continuity equation}

The external power density $\mathbb{\bar{N}}_g(\bar{n}_0, \bar{n}_1,
\bar{n}_2, \bar{n}_3)$ is defined from the total energy density
$\mathbb{\bar{W}}_g$ .
\begin{eqnarray}
\mathbb{\bar{N}}_g = \bar{v}_0 ( \mathbb{\bar{B}}_g/\bar{v}_0 +
\lozenge)^* \circ \mathbb{\bar{W}}_g
\end{eqnarray}
where, the scalar $\bar{n}_0 = \overline{\textbf{b}}^* \cdot
\overline{\textbf{w}} + \bar{v}_0 \partial_0 \bar{w}_0 + \bar{v}_0
\nabla^* \cdot \overline{\textbf{w}} $ is the power density; the
external power density $\mathbb{\bar{N}}_g$ includes the power
density $n_0$ etc. in the gravitational field.

From the above, the dot product of velocity curl with torque is the
part of power in Newtonian gravity. In some cases, comparing Eq.(26)
with Eq.(36), we find $\mathbb{\bar{C}}_g$ must equal to zero.
Otherwise, $\mathbb{\bar{N}}_g$ will change with the time.

When the quaternion coordinate system rotates, we have the external
power density $\mathbb{\bar{N}}_g' (\bar{n}'_0, \bar{n}'_1,
\bar{n}'_2, \bar{n}'_3 )$ from Eq.(2). And then, we find that the
power density will remain unchanged by Eq.(3).
\begin{eqnarray}
\bar{n}_0 = \bar{n}'_0
\end{eqnarray}

In some special cases, the $\bar{n}'_0 = 0$, and then we obtain the
energy continuity equation from Eqs.(9) and (37).
\begin{eqnarray}
\partial \bar{w}_0 / \partial r_0 + \nabla^* \cdot \overline{\textbf{w}}
+ \overline{\textbf{b}}^* \cdot \overline{\textbf{w}} / \bar{v}_0 =
0
\end{eqnarray}

If the last term is neglected, the above is reduced to
\begin{eqnarray}
\partial \bar{w}_0 / \partial r_0 + \nabla^* \cdot \overline{\textbf{w}} =
0.
\end{eqnarray}
further, if the last term is equal to zero, we have
\begin{eqnarray}
\partial \bar{w}_0 / \partial t = 0.
\nonumber
\end{eqnarray}
where, when the time $t$ is only the independent variable, the
$\partial \bar{w}_0 / \partial t$ will become the $d \bar{w}_0 / d
t$ .

The above means the strength $\overline{\textbf{b}}$ and torque
density $\overline{\textbf{w}}$ have the influence on the energy
continuity equation. And the power density, $n_0$, may be variable
under the quaternion transformation in the gravitational field.

\section{Mechanical invariants of electromagnetic field}

The octonion \cite{cayley} can be used to describe the property of
the electromagnetic field \cite{maxwell} and gravitational field,
including the scalar invariants in the case of existence of strength
and velocity curl etc.

In the case for coexistence of the electromagnetic field and the
gravitational field \cite{newton}, there are the mechanical
invariants, which include the conservation of mass, the conservation
of spin, and the conservation of energy \cite{joule}. With the
property of octonions, we find that the velocity curl and strength
have the influence on conservation laws in the electromagnetic and
gravitational fields.

From the above equations, we may obtain some kinds of mechanical
invariants with the velocity curl under the coordinate
transformation in the case for coexistence of gravitational field
and electromagnetic field. And we find that the speed of light, mass
density, spin density, and energy density are all variable in the
case for coexistence of the electromagnetic field and gravitational
field, under the octonion coordinate transformation.

\subsection{Octonion transformation}

In the octonion space, the basis vector $\mathbb{E}$ consists of the
quaternion basis vectors $\mathbb{E}_g$ and $\mathbb{E}_e$ . The
basis vector $\mathbb{E}_g = (1, \emph{\textbf{i}}_1,
\emph{\textbf{i}}_2, \emph{\textbf{i}}_3$) is the basis vector of
the quaternion space for the gravitational field, and $\mathbb{E}_e
= (\emph{\textbf{I}}_0, \emph{\textbf{I}}_1, \emph{\textbf{I}}_2,
\emph{\textbf{I}}_3$) for the electromagnetic field. While the basis
vector $\mathbb{E}_e$ is independent of the $\mathbb{E}_g$, with
$\mathbb{E}_e$ = $\mathbb{E}_g \circ \emph{\textbf{I}}_0$ .
\begin{eqnarray}
\mathbb{E} = (1, \emph{\textbf{i}}_1, \emph{\textbf{i}}_2,
\emph{\textbf{i}}_3, \emph{\textbf{I}}_0, \emph{\textbf{I}}_1,
\emph{\textbf{I}}_2, \emph{\textbf{I}}_3)
\end{eqnarray}

The octonion quantity $\mathbb{D} (d_0, d_1, d_2, d_3, D_0, D_1,
D_2, D_3 )$ is defined as follows.
\begin{eqnarray}
\mathbb{D} = d_0 + \Sigma (d_j \emph{\textbf{i}}_j) + \Sigma (D_i
\emph{\textbf{I}}_i)
\end{eqnarray}
where, $d_i$ and $D_i$ are all real; $i = 0, 1, 2, 3$; $j , k = 1,
2, 3$.

When the coordinate system is transformed into the other, the
physical quantity $\mathbb{D}$ will be transformed into the octonion
$\mathbb{D}' (d'_0 , d'_1 , d'_2 , d'_3 , D'_0 , D'_1 , D'_2 , D'_3
)$ .
\begin{equation}
\mathbb{D}' = \mathbb{K}^* \circ \mathbb{D} \circ \mathbb{K}
\end{equation}
where, $\mathbb{K}$ is the octonion, and $\mathbb{K}^* \circ
\mathbb{K} = 1$; $*$ denotes the conjugate of octonion; $\circ$ is
the octonion multiplication.

The octonion $\mathbb{D}$ satisfies the following equations.
\begin{eqnarray}
& d_0 = d'_0
\\
& \mathbb{D}^* \circ \mathbb{D} = (\mathbb{D'})^* \circ \mathbb{D'}
\end{eqnarray}

In the above equation, the scalar part $d_0$ is preserved during the
octonion coordinates are transforming. Some scalar invariants of
electromagnetic field will be obtained from the characteristics of
the octonion.

\begin{table}[t]
\caption{\label{tab:table1}The octonion multiplication table.}
\begin{ruledtabular}
\begin{tabular}{ccccccccc}
$ $ & $1$ & $\emph{\textbf{i}}_1$  & $\emph{\textbf{i}}_2$ &
$\emph{\textbf{i}}_3$  & $\emph{\textbf{I}}_0$  &
$\emph{\textbf{I}}_1$
& $\emph{\textbf{I}}_2$  & $\emph{\textbf{I}}_3$  \\
\hline $1$ & $1$ & $\emph{\textbf{i}}_1$  & $\emph{\textbf{i}}_2$ &
$\emph{\textbf{i}}_3$  & $\emph{\textbf{I}}_0$  &
$\emph{\textbf{I}}_1$
& $\emph{\textbf{I}}_2$  & $\emph{\textbf{I}}_3$  \\
$\emph{\textbf{i}}_1$ & $\emph{\textbf{i}}_1$ & $-1$ &
$\emph{\textbf{i}}_3$  & $-\emph{\textbf{i}}_2$ &
$\emph{\textbf{I}}_1$
& $-\emph{\textbf{I}}_0$ & $-\emph{\textbf{I}}_3$ & $\emph{\textbf{I}}_2$  \\
$\emph{\textbf{i}}_2$ & $\emph{\textbf{i}}_2$ &
$-\emph{\textbf{i}}_3$ & $-1$ & $\emph{\textbf{i}}_1$  &
$\emph{\textbf{I}}_2$  & $\emph{\textbf{I}}_3$
& $-\emph{\textbf{I}}_0$ & $-\emph{\textbf{I}}_1$ \\
$\emph{\textbf{i}}_3$ & $\emph{\textbf{i}}_3$ &
$\emph{\textbf{i}}_2$ & $-\emph{\textbf{i}}_1$ & $-1$ &
$\emph{\textbf{I}}_3$  & $-\emph{\textbf{I}}_2$
& $\emph{\textbf{I}}_1$  & $-\emph{\textbf{I}}_0$ \\
$\emph{\textbf{I}}_0$ & $\emph{\textbf{I}}_0$ &
$-\emph{\textbf{I}}_1$ & $-\emph{\textbf{I}}_2$ &
$-\emph{\textbf{I}}_3$ & $-1$ & $\emph{\textbf{i}}_1$
& $\emph{\textbf{i}}_2$  & $\emph{\textbf{i}}_3$  \\
$\emph{\textbf{I}}_1$ & $\emph{\textbf{I}}_1$ &
$\emph{\textbf{I}}_0$ & $-\emph{\textbf{I}}_3$ &
$\emph{\textbf{I}}_2$  & $-\emph{\textbf{i}}_1$
& $-1$ & $-\emph{\textbf{i}}_3$ & $\emph{\textbf{i}}_2$  \\
$\emph{\textbf{I}}_2$ & $\emph{\textbf{I}}_2$ &
$\emph{\textbf{I}}_3$ & $\emph{\textbf{I}}_0$  &
$-\emph{\textbf{I}}_1$ & $-\emph{\textbf{i}}_2$
& $\emph{\textbf{i}}_3$  & $-1$ & $-\emph{\textbf{i}}_1$ \\
$\emph{\textbf{I}}_3$ & $\emph{\textbf{I}}_3$ &
$-\emph{\textbf{I}}_2$ & $\emph{\textbf{I}}_1$  &
$\emph{\textbf{I}}_0$  & $-\emph{\textbf{i}}_3$
& $-\emph{\textbf{i}}_2$ & $\emph{\textbf{i}}_1$  & $-1$ \\
\end{tabular}
\end{ruledtabular}
\end{table}

\subsection{Radius vector}

In the octonion space for the gravitational and electromagnetic
fields, the radius vector $\mathbb{R} = \Sigma (r_i
\emph{\textbf{i}}_i) + \Sigma (R_i \emph{\textbf{I}}_i)$. And it can
be combined with the octonion $\mathbb{X} = \Sigma (x_i
\emph{\textbf{i}}_i) + \Sigma (X_i \emph{\textbf{I}}_i)$ to become
the compounding radius vector $\bar{\mathbb{R}}$.
\begin{eqnarray}
\mathbb{\bar{R}} = \mathbb{R} + k_{rx} \mathbb{X}
\end{eqnarray}
where, $R_0 = V_0 T$; $V_0$ represents the speed of light-like, $T$
is a time-like quantity; $X_0 = A_0 T$; $A_0$ is the scalar
potential of electromagnetic field.

In other words, the $\mathbb{\bar{R}}$ can be considered as the
radius vector in the octonion compounding space, with the basis
vector $(1, \emph{\textbf{i}}_1, \emph{\textbf{i}}_2,
\emph{\textbf{i}}_3, \emph{\textbf{I}}_0, \emph{\textbf{I}}_1,
\emph{\textbf{I}}_2, \emph{\textbf{I}}_3)$.

When the octonion coordinate system is rotated, we obtain the radius
vector $\mathbb{\bar{R}}' (\bar{r}'_0, \bar{r}'_1, \bar{r}'_2,
\bar{r}'_3, \bar{R}'_0, \bar{R}'_1, \bar{R}'_2, \bar{R}'_3 )$ from
Eqs.(42) and (45).

From Eqs.(43) and (45), we have
\begin{eqnarray}
\bar{r}_0 = \bar{r}'_0~.
\end{eqnarray}
where, $\bar{r}_i = r_i + k_{rx} x_i$ , $\bar{R}_i = R_i + k_{rx}
X_i$ .

Sometimes, the radius vector $\mathbb{\bar{R}}$ can be replaced by
the physical quantity $\mathbb{\bar{Z}} (\bar{z}_0, \bar{z}_1,
\bar{z}_2, \bar{z}_3, \bar{Z}_0, \bar{Z}_1, \bar{Z}_2, \bar{Z}_3)$,
which is defined as
\begin{eqnarray}
\mathbb{\bar{Z}} = \mathbb{\bar{R}} \circ \mathbb{\bar{R}}.
\end{eqnarray}
where, $\bar{z}_0 = (\bar{r}_0)^2 - \Sigma (\bar{r}_j)^2 - \Sigma
(\bar{R}_i)^2$.

By Eqs.(43) and (47), we have
\begin{eqnarray}
\bar{z}_0 = \bar{z}'_0~.
\end{eqnarray}

The above represents that the scalar invariant $\bar{r}_0$ and
$\bar{z}_0$ remain unchanged when the coordinate system rotates in
the octonion compounding space. And there may exist the special case
of the $x_i \neq 0$ when $r_i = R_i = 0$.

\subsection{Speed of light}

The velocity $\mathbb{V} = \Sigma (v_i \emph{\textbf{i}}_i) + \Sigma
(V_i \emph{\textbf{I}}_i)$ can be combined with the potential
$\mathbb{A} = \Sigma (a_i \emph{\textbf{i}}_i) + \Sigma (A_i
\emph{\textbf{I}}_i)$ to become the velocity $\mathbb{\bar{V}}$ in
the octonion compounding space.
\begin{eqnarray}
\mathbb{\bar{V}} = \mathbb{V} + k_{rx} \mathbb{A}
\end{eqnarray}
where, $\bar{v}_i = v_i + k_{rx} a_i$ ; $\bar{V}_i = V_i + k_{rx}
A_i$ .

In Eq.(49), the potential $\mathbb{A}$ consists of the gravitational
potential $\mathbb{A}_g = \Sigma (a_i \emph{\textbf{i}}_i)$, and the
electromagnetic potential $\mathbb{A}_e = \Sigma (A_i
\emph{\textbf{I}}_i)$ .
\begin{eqnarray}
\mathbb{A} = \mathbb{A}_g + k_{eg} \mathbb{A}_e
\end{eqnarray}
where, $k_{eg}$ is the coefficient.

When the octonion coordinate system is rotated, we have the velocity
$\mathbb{\bar{V}}' (\bar{v}'_0, \bar{v}'_1, \bar{v}'_2, \bar{v}'_3,
\bar{V}'_0, \bar{V}'_1, \bar{V}'_2, \bar{V}'_3 )$ from Eqs.(42) and
(49).

We have the scalar invariant about the speed of light in the
octonion compounding space by Eqs.(43) and (49).
\begin{eqnarray}
\bar{v}_0 = \bar{v}'_0
\end{eqnarray}

If we emphasize the definitions of radius vector Eq.(45) and
velocity Eq.(49), we obtain Galilean transformation from Eqs.(46)
and (51).
\begin{eqnarray}
\bar{r}_0 = \bar{r}'_0~, \bar{v}_0 = \bar{v}'_0~.
\end{eqnarray}

In the same way, in case of we choose definitions of physical
quantity Eq.(47) and velocity Eq.(49), we have Lorentz
transformation \cite{lorentz} from Eqs.(48) and (51).
\begin{eqnarray}
\bar{z}_0 = \bar{z}'_0~, \bar{v}_0 = \bar{v}'_0~.
\end{eqnarray}

In some special cases, the $\Sigma (\bar{R}_i \emph{\textbf{I}}_i)$
do not take part the octonion coordinate transformation, we have the
result $\Sigma (\bar{R}_i)^2 = \Sigma (\bar{R}'_i)^2$ in Eq.(48).

The above means that the speed of light, $v_0$, will be variable,
due to the existence of the scalar potential, $a_0$, of the
gravitational field. But it is not associated with the scalar
potential of electromagnetic field.

\subsection{Potential and strength}

In the octonion compounding space, the potential $\mathbb{\bar{A}} =
\Sigma (\bar{a}_i \emph{\textbf{i}}_i) + \Sigma (\bar{A}_i
\emph{\textbf{I}}_i)$ is defined from the velocity
$\mathbb{\bar{V}}$ .
\begin{eqnarray}
\mathbb{\bar{A}} = \mathbb{A} + K_{rx} \mathbb{V}
\end{eqnarray}
where, $\bar{a}_i = a_i + K_{rx} v_i$; $\bar{A}_i = A_i + K_{rx}
V_i$.

When the coordinate system is rotated, we have the potential
$\mathbb{\bar{A}}' (\bar{a}'_0, \bar{a}'_1, \bar{a}'_2, \bar{a}'_3,
\bar{A}'_0, \bar{A}'_1, \bar{A}'_2, \bar{A}'_3)$ from Eqs.(42) and
(54). And we have the invariant about the scalar potential of
gravitational field by Eqs.(43) and (54).
\begin{eqnarray}
\bar{a}_0 = \bar{a}'_0
\end{eqnarray}

In the octonion compounding space, the strength $\mathbb{\bar{B}} =
\Sigma (\bar{b}_i \emph{\textbf{i}}_i) + \Sigma (\bar{B}_i
\emph{\textbf{I}}_i)$ is defined from the potential
$\mathbb{\bar{A}}$ .
\begin{eqnarray}
\mathbb{\bar{B}} = \lozenge \circ \mathbb{\bar{A}} = \mathbb{B} +
K_{rx} \mathbb{U}
\end{eqnarray}
where, $\bar{b}_i = b_i + K_{rx} u_i$, $\bar{B}_i = B_i + K_{rx}
U_i$; the velocity $\mathbb{V} = v_0 [\lozenge \circ \mathbb{R} -
\nabla \cdot \left\{ \Sigma (r_j \emph{\textbf{i}}_j ) \right\} ]$,
the velocity curl $\mathbb{U} = \lozenge \circ \mathbb{V}$;
$\mathbb{A} = v_0 \lozenge \circ \mathbb{X}$; $\mathbb{B} = \lozenge
\circ \mathbb{A}$.

In Eq.(56), the strength $\mathbb{B} = \Sigma (b_i
\emph{\textbf{i}}_i) + \Sigma (B_i \emph{\textbf{I}}_i)$ consists of
the gravitational strength $\mathbb{B}_g$ and the electromagnetic
strength $\mathbb{B}_e$.
\begin{eqnarray}
\mathbb{B} = \lozenge \circ \mathbb{A} = \mathbb{B}_g + k_{eg}
\mathbb{B}_e
\end{eqnarray}

In the above equation, we choose the following gauge conditions to
simplify succeeding calculation.
\begin{eqnarray}
\bar{b}_0 = 0~,~ \bar{B}_0 = 0~.
\end{eqnarray}

When the coordinate system is rotated, we have the strength
$\mathbb{\bar{B}}' (\bar{b}'_0, \bar{b}'_1, \bar{b}'_2, \bar{b}'_3,
\bar{B}'_0, \bar{B}'_1, \bar{B}'_2, \bar{B}'_3)$ from Eqs.(42) and
(56). And we have the invariant about the scalar strength of
gravitational field by Eqs.(43) and (56).
\begin{eqnarray}
\bar{b}_0 = \bar{b}'_0
\end{eqnarray}

The above means that the scalar potential, $a_0$, and the scalar
strength, $b_0$, of gravitational field will be variable in the
octonion compounding space, although the $\bar{a}_0$ and the
$\bar{b}_0$ both are the scalar invariants.

\subsection{Conservation of mass}

The source density $\mathbb{\bar{S}}$ is defined from the strength
$\mathbb{\bar{B}}$ in the octonion compounding space.
\begin{eqnarray}
\mu \mathbb{\bar{S}} && = - ( \mathbb{\bar{B}}/\bar{v}_0 +
\lozenge)^* \circ \mathbb{\bar{B}}
\nonumber\\
&& = \mu_g^g \mathbb{\bar{S}}_g + k_{eg} \mu_e^g \mathbb{\bar{S}}_e
- \mathbb{\bar{B}}^* \circ \mathbb{\bar{B}}/\bar{v}_0
\end{eqnarray}
where, $k_{eg}^2 = \mu_g^g /\mu_e^g$; $q$ is the electric charge
density; $\mu_e^g$ is the constant; the velocity $\mathbb{\bar{V}}_e
= \Sigma (\bar{V}_i \emph{\textbf{I}}_i) $; the electric current
density $\mathbb{\bar{S}}_e = q \mathbb{\bar{V}}_e$ is the source of
electromagnetic field.

In some cases, the electric charge is combined with the mass to
become the electron or proton etc., we have the condition $
\bar{R}_i \emph{\textbf{I}}_i = \bar{r}_i \emph{\textbf{i}}_i \circ
\emph{\textbf{I}}_0$ and $\bar{V}_i \emph{\textbf{I}}_i = \bar{v}_i
\emph{\textbf{i}}_i \circ \emph{\textbf{I}}_0$ .

The $\mathbb{\bar{B}}^* \circ \mathbb{\bar{B}}/(2\mu_g^g)$ is the
energy density, and includes that of the electromagnetic field.
\begin{eqnarray}
\mathbb{\bar{B}}^* \circ \mathbb{\bar{B}}/ \mu_g^g =
\mathbb{\bar{B}}_g^* \circ \mathbb{\bar{B}}_g / \mu_g^g +
\mathbb{\bar{B}}_e^* \circ \mathbb{\bar{B}}_e / \mu_e^g
\end{eqnarray}

The linear momentum density $\mathbb{\bar{P}} = \mu \mathbb{\bar{S}}
/ \mu_g^g$ is written as
\begin{eqnarray}
\mathbb{\bar{P}} = \widehat{m} \bar{v}_0 + \Sigma (m \bar{v}_j
\emph{\textbf{i}}_j ) + \Sigma (M \bar{V}_i \emph{\textbf{i}}_i
\circ \emph{\textbf{I}}_0 ).
\end{eqnarray}
where, $\widehat{m} = m - (\mathbb{\bar{B}}^* \circ \mathbb{\bar{B}}
/ \mu_g^g )/\bar{v}_0^2 $; $M = k_{eg} \mu_e^g q / \mu_g^g$ .

The above means that the gravitational mass density $\widehat{m}$ is
changed with either the strength or the velocity curl in the
electromagnetic and gravitational fields.

From Eq.(42), we have the linear momentum density,
$\mathbb{\bar{P}}' (\widehat{m}' \bar{v}'_0, m' \bar{v}'_1, m'
\bar{v}'_2, m' \bar{v}'_3, M' \bar{V}'_0, M' \bar{V}'_1, M'
\bar{V}'_2, M' \bar{V}'_3 )$, when the coordinate system is rotated.
And we obtain the invariant equation from Eqs.(43) and (62).
\begin{eqnarray}
\widehat{m} \bar{v}_0 = \widehat{m}' \bar{v}'_0
\end{eqnarray}

Under Eqs.(51) and (63), we find that the gravitational mass density
$\widehat{m}$ remains unchanged. And then we have the conservation
of mass.
\begin{eqnarray}
\widehat{m} = \widehat{m}'
\end{eqnarray}

The above means that if we choose the definitions of velocity and
linear momentum, the inertial mass density and the gravitational
mass density will keep unchanged, under the octonion coordinate
transformation in Eq.(42) in the octonion compounding space. The
results are the same as that in the octonion space in the
electromagnetic field and gravitational field.

\subsection{Mass continuity equation}

In the octonion compounding space, the applied force density
$\mathbb{\bar{F}} = \Sigma (\bar{f}_i \emph{\textbf{i}}_i ) + \Sigma
(\bar{F}_i \emph{\textbf{I}}_i )$ is defined from the linear
momentum density $\mathbb{\bar{P}}$ in Eq.(62).
\begin{eqnarray}
\mathbb{\bar{F}} = \bar{v}_0 (\mathbb{\bar{B}}/\bar{v}_0 + \lozenge
)^* \circ \mathbb{\bar{P}}
\end{eqnarray}
where, $\bar{f}_0 = \bar{v}_0 \partial \bar{p}_0 / \partial r_0 +
\bar{v}_0 \Sigma ( \partial \bar{p}_j / \partial r_j ) + \Sigma (
\bar{b}_j \bar{p}_j + \bar{B}_j \bar{P}_j ) $; $\bar{p}_0 =
\widehat{m} \bar{v}_0$, $\bar{p}_j = m \bar{v}_j $; $\bar{P}_i = M
\bar{V}_i $; and the applied force includes the gravity, the
inertial force, the Lorentz force, and the interacting force between
magnetic strength with magnetic moment, etc.

When the coordinate system rotates, we have the force density
$\mathbb{\bar{F}}' (\bar{f}'_0, \bar{f}'_1, \bar{f}'_2, \bar{f}'_3,
\bar{F}'_0, \bar{F}'_1, \bar{F}'_2, \bar{F}'_3)$.

By Eq.(43), we have
\begin{eqnarray}
\bar{f}_0 = \bar{f}'_0~.
\end{eqnarray}

When $\bar{f}'_0 = 0$ in the above, we have the mass continuity
equation in the case for coexistence of the gravitational field and
electromagnetic field.
\begin{eqnarray}
\partial \bar{p}_0 / \partial r_0 + \Sigma ( \partial \bar{p}_j /
\partial r_j ) + \Sigma ( \bar{b}_j \bar{p}_j + \bar{B}_j \bar{P}_j ) / \bar{v}_0 = 0
\end{eqnarray}

If the $a_i = A_i = 0$ and $\bar{b}_i = \bar{B}_i = 0$, the above
will be reduced to the following equation.
\begin{eqnarray}
\partial m / \partial t + \Sigma (  \partial p_j /
\partial r_j ) = 0
\end{eqnarray}

The above states that the potential, the strength, and the velocity
curl have the small influence on the mass continuity equation in the
gravitational field and electromagnetic field, although the $\Sigma
( \bar{b}_j \bar{p}_j + \bar{B}_j \bar{P}_j ) / \bar{v}_0$ and
$\triangle m$ both are usually very tiny when the fields are weak.
In case of we choose the definitions of the applied force and
velocity in the octonion compounding space, the mass continuity
equation will be the invariant equation under the octonion
transformation in Eq.(42).

\subsection{$\bar{f}_0$ continuity equation}

In the octonion compounding space, the new density $\mathbb{\bar{C}}
= \Sigma (\bar{c}_i \emph{\textbf{i}}_i ) + \Sigma (\bar{C}_i
\emph{\textbf{I}}_i )$ is defined from the applied force density
$\mathbb{\bar{F}}$ in Eq.(65).
\begin{eqnarray}
\mathbb{\bar{C}} = \bar{v}_0 (\mathbb{\bar{B}}/\bar{v}_0 + \lozenge
)^* \circ \mathbb{\bar{F}}
\end{eqnarray}
where, $\bar{c}_0 = \bar{v}_0 \partial \bar{f}_0 / \partial r_0 +
\bar{v}_0 \Sigma ( \partial \bar{f}_j / \partial r_j ) + \Sigma (
\bar{b}_j \bar{f}_j + \bar{B}_j \bar{F}_j ) $.

When the octonion coordinate system rotates, we have the density
$\mathbb{\bar{C}}' (\bar{c}'_0, \bar{c}'_1, \bar{c}'_2, \bar{c}'_3,
\bar{C}'_0, \bar{C}'_1, \bar{C}'_2, \bar{C}'_3)$.

By Eq.(43) and the above, we have
\begin{eqnarray}
\bar{c}_0 = \bar{c}'_0~.
\end{eqnarray}

If $\bar{c}'_0 = 0$ in the above, we have the continuity equation
about $\bar{f}_0$ in the case for coexistence of gravitational field
and electromagnetic field.
\begin{eqnarray}
\partial \bar{f}_0 / \partial r_0 + \Sigma ( \partial \bar{f}_j /
\partial r_j ) + \Sigma ( \bar{b}_j \bar{f}_j + \bar{B}_j \bar{F}_j ) / \bar{v}_0 = 0
\end{eqnarray}

When the $a_i = A_i = 0$ and $\bar{b}_i = \bar{B}_i = 0$, the above
will be reduced to the following equation.
\begin{eqnarray}
\partial \bar{f}_0 / \partial r_0 + \Sigma ( \partial \bar{f}_j /
\partial r_j ) = 0
\end{eqnarray}

The above states that the potential, the strength, and the velocity
curl have small influence on the continuity equation about
$\bar{f}_0$, although the $\Sigma ( \bar{b}_j \bar{f}_j + \bar{B}_j
\bar{F}_j ) / \bar{v}_0$ is usually very tiny when the fields are
weak. When we choose definitions of $\mathbb{\bar{V}}$ and
$\mathbb{\bar{C}}$, this continuity equation will be the invariant
equation under the octonion coordinate transformation.

\subsection{Spin density}

The angular momentum density $\mathbb{\bar{L}} = \Sigma (\bar{l}_i
\emph{\textbf{i}}_i ) + \Sigma (\bar{L}_i \emph{\textbf{I}}_i )$ is
defined from the radius vector $\mathbb{\bar{R}}$ and linear
momentum density $\mathbb{\bar{P}}$ in the octonion compounding
space.
\begin{eqnarray}
\mathbb{\bar{L}} = \mathbb{\bar{R}} \circ \mathbb{\bar{P}}
\end{eqnarray}
where, the scalar $\bar{l}_0 = \bar{r}_0 \bar{p}_0 - \Sigma
(\bar{r}_j \bar{p}_j ) - \Sigma (\bar{R}_i \bar{P}_i )$.

The $\bar{l}_0$ is considered as the spin angular momentum density
in the octonion compounding space. The angular momentum includes the
orbital angular momentum and spin angular momentum in the
gravitational field and the electromagnetic field.

When the octonion coordinate system is rotated, we have the angular
momentum density $\mathbb{\bar{L}}' = \Sigma ( \bar{l}'_i
\emph{\textbf{i}}'_i + \bar{L}'_i \emph{\textbf{I}}'_i )$. Under the
octonion coordinate transformation, the spin density remains
unchanged from Eq.(43). In other words, we have the conservation of
spin as follows.
\begin{eqnarray}
\bar{l}_0 = \bar{l}'_0
\end{eqnarray}

The above means the velocity, velocity curl, potential, and strength
have the influence on the orbital angular momentum and spin angular
momentum. Meanwhile the spin angular momentum density $l_0$ will be
variable in the gravitational field and electromagnetic field,
although the $\bar{l}_0$ is invariable under the octonion
transformation.

\begin{table}[b]
\caption{\label{tab:table1}The definitions of mechanics invariants
of gravitational field and electromagnetic field in the octonion
compounding space.}
\begin{ruledtabular}
\begin{tabular}{lll}
$definition$          &    $invariant $                &    $ meaning $                            \\
\hline
$\mathbb{\bar{R}}$    &    $\bar{r}_0 = \bar{r}'_0 $   &    $Galilean~invariant$                   \\
$\mathbb{\mathbb{\bar{R}}} \circ \mathbb{\bar{R}}$     &    $\bar{z}_0 = \bar{z}'_0 $  &  $Lorentz~invariant$  \\
$\mathbb{\bar{V}}$    &    $\bar{v}_0 = \bar{v}'_0$    &    $invariable~speed~of~light$            \\
$\mathbb{\bar{A}}$    &    $\bar{a}_0 = \bar{a}'_0$    &    $invariable~scalar~potential$          \\
$\mathbb{\bar{B}}$    &    $\bar{b}_0 = \bar{b}'_0$    &    $invariable~gauge$                     \\
$\mathbb{\bar{P}}$    &    $\bar{p}_0 = \bar{p}'_0$    &    $conservation~of~mass$                 \\
$\mathbb{\bar{F}}$    &    $\bar{f}_0 = \bar{f}'_0$    &    $mass~continuity~equation$             \\
$\mathbb{\bar{L}}$    &    $\bar{l}_0 = \bar{l}'_0$    &    $invariable~spin~density$              \\
$\mathbb{\bar{W}}$    &    $\bar{w}_0 = \bar{w}'_0$    &    $conservation~of~energy$               \\
$\mathbb{\bar{N}}$    &    $\bar{n}_0 = \bar{n}'_0$    &    $energy~continuity~equation$           \\
\end{tabular}
\end{ruledtabular}
\end{table}

\subsection{Conservation of energy}

The total energy density $\mathbb{\bar{W}} = \Sigma (\bar{w}_i
\emph{\textbf{i}}_i ) + \Sigma (\bar{W}_i \emph{\textbf{I}}_i )$ is
defined from the angular momentum density $\mathbb{\bar{L}}$.
\begin{eqnarray}
\mathbb{\bar{W}} = \bar{v}_0 ( \mathbb{\bar{B}}/\bar{v}_0 +
\lozenge) \circ \mathbb{\bar{L}}
\end{eqnarray}
where, the scalar $\bar{w}_0 = \bar{v}_0 \partial_0 \bar{l}_0 +
(\bar{v}_0 \nabla + \overline{\textbf{h}}) \cdot
\overline{\textbf{j}} + \overline{\textbf{H}} \cdot
\overline{\textbf{J}}$; $\overline{\textbf{h}} = \Sigma (\bar{b}_j
\emph{\textbf{i}}_j)$; $\overline{\textbf{H}} = \Sigma (\bar{B}_j
\emph{\textbf{I}}_j)$; $\overline{\textbf{j}} = \Sigma (\bar{l}_j
\emph{\textbf{i}}_j)$; $\overline{\textbf{J}} = \Sigma (\bar{L}_j
\emph{\textbf{I}}_j)$; the total energy includes the potential
energy, the kinetic energy, the torque, and the work, etc. in the
gravitational field and the electromagnetic field \cite{heaviside}.

When the coordinate system is rotated, we have the energy density
$\mathbb{\bar{W}}' = \Sigma ( \bar{w}'_i \emph{\textbf{i}}'_i +
\bar{W}'_i \emph{\textbf{I}}'_i )$. Under the octonion
transformation, the scalar part of total energy density is the
energy density and remains unchanged \cite{einstein}. So we have the
conservation of energy as follows.
\begin{eqnarray}
\bar{w}_0 = \bar{w}'_0
\end{eqnarray}

In some special cases, the right side is equal to zero. We obtain
the spin continuity equation.
\begin{eqnarray}
\partial \bar{l}_0 / \partial r_0 + \nabla \cdot \overline{\textbf{j}}
+ (\overline{\textbf{h}} \cdot \overline{\textbf{j}} +
\overline{\textbf{H}} \cdot \overline{\textbf{J}}) / \bar{v}_0 = 0
\end{eqnarray}

If the last term is neglected, the above is reduced to
\begin{eqnarray}
\partial \bar{l}_0 / \partial r_0 + \nabla \cdot \overline{\textbf{j}} = 0
\end{eqnarray}
further, if the last term is equal to zero, we have
\begin{eqnarray}
\partial \bar{l}_0 / \partial t = 0 .
\end{eqnarray}
where, when the time $t$ is only the independent variable, the
$\partial \bar{l}_0 / \partial t$ will become the $d \bar{l}_0 / d
t$ .

The above means the energy density $w_0$ is variable in the case for
coexistence of the gravitational field and the electromagnetic
field, because the velocity, velocity curl, potential, and strength
have the influence on the angular momentum density. While the scalar
$\bar{w}_0$ is the invariant under the octonion coordinate
transformation from Eqs.(43) and (76).

\subsection{Energy continuity equation}

In the octonion compounding space with the velocity curl, the
external power density $\mathbb{\bar{N}}$  can be defined from the
total energy density $\mathbb{\bar{W}}$ in Eq.(75).
\begin{eqnarray}
\mathbb{\bar{N}} = \bar{v}_0 ( \mathbb{\bar{B}}/\bar{v}_0 +
\lozenge)^* \circ \mathbb{\bar{W}}
\end{eqnarray}
where, the external power density $\mathbb{\bar{N}}$ includes the
power density in the gravitational and electromagnetic fields.

The external power density can be rewritten as follows.
\begin{eqnarray}
\mathbb{\bar{N}} = \bar{n}_0 + \Sigma (\bar{n}_j \emph{\textbf{i}}_j
) + \Sigma (\bar{N}_i \emph{\textbf{I}}_i )
\end{eqnarray}
where, the scalar $\bar{n}_0 = \bar{v}_0 \partial_0 \bar{w}_0 +
(\bar{v}_0 \nabla + \overline{\textbf{h}})^* \cdot
\overline{\textbf{y}} + \overline{\textbf{H}}^* \cdot
\overline{\textbf{Y}}$; $\overline{\textbf{y}} = \Sigma (\bar{w}_j
\emph{\textbf{i}}_j)$; $\overline{\textbf{Y}} = \Sigma (\bar{W}_j
\emph{\textbf{I}}_j)$.

When the coordinate system is rotated, we have the external power
density $\mathbb{\bar{N}}' = \Sigma ( \bar{n}'_i
\emph{\textbf{i}}'_i + \bar{N}'_i \emph{\textbf{I}}'_i )$. Under the
octonion coordinate transformation, the scalar part of external
power density is the power density and remains unchanged by Eq.(43).
That is the conservation of power.
\begin{eqnarray}
\bar{n}_0 = \bar{n}'_0
\end{eqnarray}

In the special cases, the right side is equal to zero. And then, we
obtain the energy continuity equation.
\begin{eqnarray}
\partial_0 \bar{w}_0 + \nabla^* \cdot \overline{\textbf{y}} +
(\overline{\textbf{h}}^* \cdot \overline{\textbf{y}} +
\overline{\textbf{H}}^* \cdot \overline{\textbf{Y}}) / \bar{v}_0 = 0
\end{eqnarray}

If the last term is neglected, the above is reduced to
\begin{eqnarray}
\partial_0 \bar{w}_0 + \nabla^* \cdot \overline{\textbf{y}} = 0
\end{eqnarray}
further, if the last term is equal to zero, we have
\begin{eqnarray}
\partial \bar{w}_0 / \partial t = 0.
\end{eqnarray}
where, when the time $t$ is only the independent variable, the
$\partial \bar{w}_0 / \partial t$ will become the $d \bar{w}_0 / d
t$ .

The above means that the power density $n_0$ will be variable in the
case for coexistence of the gravitational field and electromagnetic
field, although the $\bar{n}_0$ is the scalar invariant under the
octonion transformation. And the potential, strength, velocity curl,
or torque density, etc. have the influence on the energy continuity
equation in the gravitational field and the electromagnetic field.

\section{Electric invariants of electromagnetic field}

In the octonion space for the electromagnetic field and
gravitational field, there exist the electric invariants, which
include the conservation of charge, conservation of spin magnetic
moment, charge continuity equation, and conservation of energy-like,
etc.

The electric invariants of electromagnetic field can be illustrated
by octonions in the case for coexistence of the electromagnetic
field and gravitational field.

With the property of the algebra of octonions, we find that the
velocity curl and strength have the influence on the electric
invariants in the electromagnetic field and gravitational field with
the strength and velocity curl.

\subsection{Radius vector}

In the octonion compounding space, one new octonion quantity
$\mathbb{\bar{R}}_q = \mathbb{\bar{R}} \circ \emph{\textbf{I}}_0^*$
can be defined from Eq.(45).
\begin{eqnarray}
\mathbb{\bar{R}}_q = \Sigma (\bar{R}_i \emph{\textbf{i}}_i) - \Sigma
(\bar{r}_i \emph{\textbf{I}}_i)
\end{eqnarray}

When the octonion coordinate system is rotated, we obtain the radius
vector $\mathbb{\bar{R}}' (\bar{r}'_0, \bar{r}'_1, \bar{r}'_2,
\bar{r}'_3, \bar{R}'_0, \bar{R}'_1, \bar{R}'_2, \bar{R}'_3 )$ from
Eqs.(42) and (86).

From Eqs.(43) and (86), we have
\begin{eqnarray}
\bar{R}_0 = \bar{R}'_0~.
\end{eqnarray}

Further, the radius vector $\mathbb{\bar{R}}$ can be replaced by the
physical quantity $\mathbb{\bar{Z}} (\bar{z}_0, \bar{z}_1,
\bar{z}_2, \bar{z}_3, \bar{Z}_0, \bar{Z}_1, \bar{Z}_2, \bar{Z}_3)$,
which is defined as
\begin{eqnarray}
\mathbb{\bar{Z}} = \mathbb{\bar{R}} \circ \mathbb{\bar{R}}.
\end{eqnarray}
where, $\bar{Z}_0 = 2 \bar{r}_0 \bar{R}_0$ .

Similarly, the new octonion quantity $\mathbb{\bar{Z}}_q =
\mathbb{\bar{Z}} \circ \emph{\textbf{I}}_0^*$ can be defined from
the above.
\begin{eqnarray}
\mathbb{\bar{Z}}_q = \Sigma (\bar{Z}_i \emph{\textbf{i}}_i) - \Sigma
(\bar{z}_i \emph{\textbf{I}}_i)
\end{eqnarray}

By Eqs.(43) and (89), we have
\begin{eqnarray}
\bar{Z}_0 = \bar{Z}'_0~.
\end{eqnarray}

The above represents that the scalar invariant $\bar{R}_0$ and
$\bar{Z}_0$ remain unchanged when the coordinate system rotates in
the octonion compounding space.

\subsection{Speed of light-like}

In the octonion compounding space, one new octonion quantity
$\mathbb{\bar{V}}_q = \mathbb{\bar{V}} \circ \emph{\textbf{I}}_0^*$
can be defined from Eq.(49).
\begin{eqnarray}
\mathbb{\bar{V}}_q = \Sigma (\bar{V}_i \emph{\textbf{i}}_i) - \Sigma
(\bar{v}_i \emph{\textbf{I}}_i)
\end{eqnarray}

When the octonion coordinate system is rotated, we have the velocity
$\mathbb{\bar{V}}' (\bar{v}'_0, \bar{v}'_1, \bar{v}'_2, \bar{v}'_3,
\bar{V}'_0, \bar{V}'_1, \bar{V}'_2, \bar{V}'_3 )$ from Eqs.(42) and
(49). And we have the scalar invariant about the speed of light in
the octonion compounding space from the above.
\begin{eqnarray}
\bar{V}_0 = \bar{V}'_0
\end{eqnarray}

If we choose the definition of radius vector Eq.(45) and velocity
Eq.(49), we obtain Galilean transformation from Eqs.(87) and (92).
\begin{eqnarray}
\bar{R}_0 = \bar{R}'_0~, \bar{V}_0 = \bar{V}'_0~.
\end{eqnarray}

Furthermore, if we choose the definitions of the radius vector
Eq.(88) and velocity Eq.(49), we shall obtain one new transformation
rather than Lorentz transformation from Eqs.(90) and (92).
\begin{eqnarray}
\bar{Z}_0 = \bar{Z}'_0~, \bar{V}_0 = \bar{V}'_0~.
\end{eqnarray}

The above means that the speed of light-like, $V_0$, will be
variable, due to the existence of the scalar potential, $A_0$, of
the electromagnetic field. In some special cases, if $\bar{R}_0 =
\bar{r}_0$ , Eq.(94) will be reduced to Eq.(93).

\subsection{Potential and strength}

In the octonion compounding space, one new octonion quantity
$\mathbb{\bar{A}}_q = \mathbb{\bar{A}} \circ \emph{\textbf{I}}_0^*$
can be defined from Eq.(54).
\begin{eqnarray}
\mathbb{\bar{A}}_q = \Sigma (\bar{A}_i \emph{\textbf{i}}_i) - \Sigma
(\bar{a}_i \emph{\textbf{I}}_i)
\end{eqnarray}

When the coordinate system is rotated, we obtain the potential
$\mathbb{\bar{A}}' (\bar{a}'_0, \bar{a}'_1, \bar{a}'_2, \bar{a}'_3,
\bar{A}'_0, \bar{A}'_1, \bar{A}'_2, \bar{A}'_3)$ from Eqs.(42) and
(54). And we have the invariant about the scalar potential of
electromagnetic field by Eqs.(43) and (95).
\begin{eqnarray}
\bar{A}_0 = \bar{A}'_0
\end{eqnarray}

In the octonion compounding space, one new octonion quantity
$\mathbb{\bar{B}}_q = \mathbb{\bar{B}} \circ \emph{\textbf{I}}_0^*$
can be defined from Eq.(56).
\begin{eqnarray}
\mathbb{\bar{B}}_q = \Sigma (\bar{B}_i \emph{\textbf{i}}_i) - \Sigma
(\bar{b}_i \emph{\textbf{I}}_i)
\end{eqnarray}

When the coordinate system is rotated, we have the strength
$\mathbb{\bar{B}}' (\bar{b}'_0, \bar{b}'_1, \bar{b}'_2, \bar{b}'_3,
\bar{B}'_0, \bar{B}'_1, \bar{B}'_2, \bar{B}'_3)$ from Eqs.(42) and
(56). And we have the invariant about the scalar strength of
electromagnetic field by Eqs.(43) and (97).
\begin{eqnarray}
\bar{B}_0 = \bar{B}'_0
\end{eqnarray}

The above means that the scalar potential, $A_0$, and the scalar
strength, $B_0$, of electromagnetic field will be variable in the
octonion compounding space, although the $\bar{A}_0$ and the
$\bar{B}_0$ both are the scalar invariants.

\subsection{Conservation of charge}

In the octonion compounding space, one new octonion quantity
$\mathbb{\bar{P}}_q = \mathbb{\bar{P}} \circ \emph{\textbf{I}}_0^*$
can be defined from Eq.(62).
\begin{eqnarray}
\mathbb{\bar{P}}_q = \Sigma (\bar{P}_i \emph{\textbf{i}}_i) - \Sigma
(\bar{p}_i \emph{\textbf{I}}_i)
\end{eqnarray}

From Eq.(42), we have the linear momentum density,
$\mathbb{\bar{P}}' (\widehat{m}' \bar{v}'_0, m' \bar{v}'_1, m'
\bar{v}'_2, m' \bar{v}'_3, M' \bar{V}'_0, M' \bar{V}'_1, M'
\bar{V}'_2, M' \bar{V}'_3 )$, when the coordinate system is rotated.
Thus we obtain the invariant equation from Eqs.(43) and (99). And
the scalar $M$ is about the electric charge density $q$ .
\begin{eqnarray}
M \bar{V}_0 = M' \bar{V}'_0
\end{eqnarray}

Under Eqs.(92) and (100), we have the conservation of charge as
follows.
\begin{eqnarray}
M = M'
\end{eqnarray}

The above means that if we choose the definition of the velocity and
linear momentum, the charge density will keep unchanged, under the
coordinate transformation in Eq.(42) in the octonion compounding
space.

\subsection{Charge continuity equation}

In the octonion compounding space, one new octonion quantity
$\mathbb{\bar{F}}_q = \mathbb{\bar{F}} \circ \emph{\textbf{I}}_0^*$
can be defined from Eq.(65).
\begin{eqnarray}
\mathbb{\bar{F}}_q = \Sigma (\bar{F}_i \emph{\textbf{i}}_i) - \Sigma
(\bar{f}_i \emph{\textbf{I}}_i)
\end{eqnarray}

When the coordinate system rotates, we have the force density
$\mathbb{\bar{F}}' (\bar{f}'_0, \bar{f}'_1, \bar{f}'_2, \bar{f}'_3,
\bar{F}'_0, \bar{F}'_1, \bar{F}'_2, \bar{F}'_3)$.

By Eq.(43) and the above, we have
\begin{eqnarray}
\bar{F}_0 = \bar{F}'_0
\end{eqnarray}

When the right side is zero in the above, we have the charge
continuity equation in the case for coexistence of the gravitational
field and electromagnetic field.
\begin{eqnarray}
\partial \bar{P}_0 / \partial r_0 + \Sigma ( \partial \bar{P}_j /
\partial r_j ) + \Sigma ( \bar{b}_j \bar{P}_j - \bar{B}_j \bar{p}_j ) / \bar{v}_0 = 0
\end{eqnarray}

If the $a_i = A_i = 0$ and $\bar{b}_i = \bar{B}_i = 0$, the above
will be reduced to the following equation.
\begin{eqnarray}
\partial M / \partial t + \Sigma (  \partial P_j /
\partial r_j ) = 0
\end{eqnarray}

The above states that the potential, the strength, and the velocity
curl have the small influence on the charge continuity equation in
the gravitational field and electromagnetic field, although the
$\Sigma ( \bar{b}_j \bar{P}_j - \bar{B}_j \bar{p}_j ) / \bar{v}_0$
and $\triangle m$ both are usually very tiny when the fields are
weak.

\subsection{$\bar{F}_0$ continuity equation}

In the octonion compounding space, one new octonion quantity
$\mathbb{\bar{C}}_q = \mathbb{\bar{C}} \circ \emph{\textbf{I}}_0^*$
can be defined from Eq.(69).
\begin{eqnarray}
\mathbb{\bar{C}}_q = \Sigma (\bar{C}_i \emph{\textbf{i}}_i) - \Sigma
(\bar{c}_i \emph{\textbf{I}}_i)
\end{eqnarray}

When the octonion coordinate system rotates, we have the density
$\mathbb{\bar{C}}' (\bar{c}'_0, \bar{c}'_1, \bar{c}'_2, \bar{c}'_3,
\bar{C}'_0, \bar{C}'_1, \bar{C}'_2, \bar{C}'_3)$.

By Eq.(43) and the above, we have
\begin{eqnarray}
\bar{C}_0 = \bar{C}'_0~.
\end{eqnarray}

When the right side is zero in the above, we have the continuity
equation about $\bar{F}_0$ in the case for coexistence of the
gravitational field and electromagnetic field.
\begin{eqnarray}
\partial \bar{F}_0 / \partial r_0 + \Sigma ( \partial \bar{F}_j /
\partial r_j ) + \Sigma ( \bar{b}_j \bar{F}_j - \bar{B}_j \bar{f}_j ) / \bar{v}_0 = 0
\end{eqnarray}

If the $a_i = A_i = 0$ and $\bar{b}_i = \bar{B}_i = 0$, the above
will be reduced to the following equation.
\begin{eqnarray}
\partial \bar{F}_0 / \partial r_0 + \Sigma ( \partial \bar{F}_j /
\partial r_j ) = 0
\end{eqnarray}

The above states that the potential, the strength, and the velocity
curl have the small influence on the continuity equation about
$\bar{F}_0$ in the gravitational field and electromagnetic field,
although the $\Sigma ( \bar{b}_j \bar{F}_j - \bar{B}_j \bar{f}_j ) /
\bar{v}_0$ is usually very tiny when the fields are weak.

\subsection{Spin magnetic moment}

In the octonion compounding space, one new octonion quantity
$\mathbb{\bar{L}}_q = \mathbb{\bar{L}} \circ \emph{\textbf{I}}_0^*$
can be defined from Eq.(73).
\begin{eqnarray}
\mathbb{\bar{L}}_q = \Sigma (\bar{L}_i \emph{\textbf{i}}_i) - \Sigma
(\bar{l}_i \emph{\textbf{I}}_i)
\end{eqnarray}

When the coordinate system rotates, we have the angular momentum
density $\mathbb{\bar{L}}' = \Sigma ( \bar{l}'_i
\emph{\textbf{i}}'_i + \bar{L}'_i \emph{\textbf{I}}'_i )$. Under the
coordinate transformation, the scalar part $\bar{L}_0$ about the
spin magnetic moment density will remain unchanged from Eqs.(43) and
(110). And we have the conservation of spin magnetic moment.
\begin{eqnarray}
\bar{L}_0 = \bar{L}'_0
\end{eqnarray}

The above means the velocity, velocity curl, potential, and strength
have the influence on the angular momentum, including the spin
magnetic moment density. While the scalar $L_0$ about the spin
magnetic moment density will be variable in the gravitational and
electromagnetic fields, although the $\bar{L}_0$ is invariable under
the octonion coordinate transformation.

\begin{table}[b]
\caption{\label{tab:table1}The definitions of electric invariants of
gravitational field and electromagnetic field in the octonion
compounding space.}
\begin{ruledtabular}
\begin{tabular}{lll}
$definition$            &    $invariant $                &    $ meaning $                            \\
\hline
$\mathbb{\bar{R}}_q$    &    $\bar{R}_0 = \bar{R}'_0 $   &    $Galilean~invariant$                   \\
$\mathbb{\mathbb{\bar{R}}}_q \circ \mathbb{\bar{R}}_q$   &    $\bar{Z}_0 = \bar{Z}'_0 $  &  $new~invariant$  \\
$\mathbb{\bar{V}}_q$    &    $\bar{V}_0 = \bar{V}'_0$    &    $speed~of~light-like$                  \\
$\mathbb{\bar{A}}_q$    &    $\bar{A}_0 = \bar{A}'_0$    &    $scalar~potential$                     \\
$\mathbb{\bar{B}}_q$    &    $\bar{B}_0 = \bar{B}'_0$    &    $gauge~equation$                       \\
$\mathbb{\bar{P}}_q$    &    $\bar{P}_0 = \bar{P}'_0$    &    $conservation~of~charge$               \\
$\mathbb{\bar{F}}_q$    &    $\bar{F}_0 = \bar{F}'_0$    &    $charge~continuity~equation$           \\
$\mathbb{\bar{L}}_q$    &    $\bar{L}_0 = \bar{L}'_0$    &    $spin~magnetic~moment~density$         \\
$\mathbb{\bar{W}}_q$    &    $\bar{W}_0 = \bar{W}'_0$    &    $conservation~of~energy-like$          \\
$\mathbb{\bar{N}}_q$    &    $\bar{N}_0 = \bar{N}'_0$    &    $conservation~of~power-like$           \\
\end{tabular}
\end{ruledtabular}
\end{table}

\subsection{Conservation of energy-like}

In the octonion compounding space, one new octonion quantity
$\mathbb{\bar{W}}_q = \mathbb{\bar{W}} \circ \emph{\textbf{I}}_0^*$
can be defined from Eq.(75).
\begin{eqnarray}
\mathbb{\bar{W}}_q = \Sigma (\bar{W}_i \emph{\textbf{i}}_i) - \Sigma
(\bar{w}_i \emph{\textbf{I}}_i)
\end{eqnarray}

When the coordinate system is rotated, we have the energy density
$\mathbb{\bar{W}}' = \Sigma ( \bar{w}'_i \emph{\textbf{i}}'_i +
\bar{W}'_i \emph{\textbf{I}}'_i )$. Under the octonion
transformation, the scalar part $\bar{W}_0$ remains unchanged by
Eqs.(43) and (75). And then we have the conservation of energy-like
as follows.
\begin{eqnarray}
\bar{W}_0 = \bar{W}'_0
\end{eqnarray}

In some special cases, if $\bar{W}'_0 = 0$ in the above, we obtain
the continuity equation of spin magnetic moment.
\begin{eqnarray}
\partial \bar{L}_0 / \partial r_0 + \Sigma (
\partial \bar{L}_j / \partial r_j ) + \Sigma ( \bar{b}_j \bar{L}_j -
\bar{B}_j \bar{l}_j )/\bar{v}_0 = 0
\nonumber
\end{eqnarray}

If the last term is neglected, the above is reduced to,
\begin{eqnarray}
\partial \bar{L}_0 / \partial r_0 + \Sigma (
\partial \bar{L}_j / \partial r_j ) = 0
\end{eqnarray}
further, if the last term is equal to zero, we have
\begin{eqnarray}
\partial \bar{L}_0 / \partial t = 0 .
\end{eqnarray}
where, when the time $t$ is only the independent variable, the
$\partial \bar{L}_0 / \partial t$ will become the $d \bar{L}_0 / d
t$ .

The above means that the energy-like density $W_0$ is variable in
the case for coexistence of the gravitational field and the
electromagnetic field, because the velocity, velocity curl,
potential, and strength have the influence on the spin magnetic
moment density. While the scalar $\bar{W}_0$ is the invariant under
the octonion transformation from Eqs.(43) and (113).

\subsection{Energy-like continuity equation}

In the octonion compounding space, one new octonion quantity
$\mathbb{\bar{N}}_q = \mathbb{\bar{N}} \circ \emph{\textbf{I}}_0^*$
can be defined from Eq.(80).
\begin{eqnarray}
\mathbb{\bar{N}}_q = \Sigma (\bar{N}_i \emph{\textbf{i}}_i) - \Sigma
(\bar{n}_i \emph{\textbf{I}}_i)
\end{eqnarray}

When the octonion coordinate system rotates, we have the external
power density $\mathbb{\bar{N}}' = \Sigma ( \bar{n}'_i
\emph{\textbf{i}}'_i + \bar{N}'_i \emph{\textbf{I}}'_i )$. Under the
octonion coordinate transformation, the scalar part $\bar{N}_0$ will
remain unchanged by Eqs.(43) and (116). And then we have
conservation of power-like.
\begin{eqnarray}
\bar{N}_0 = \bar{N}'_0
\end{eqnarray}

In a special case, the right side is equal to zero. And then, we
obtain the energy-like continuity equation.
\begin{eqnarray}
\partial \bar{W}_0 / \partial r_0 + \Sigma ( \partial \bar{W}_j /
\partial r_j ) + \Sigma ( \bar{b}_j \bar{W}_j - \bar{B}_j \bar{w}_j
)/\bar{v}_0 = 0
\nonumber
\end{eqnarray}

If the last term is neglected, the above is reduced to,
\begin{eqnarray}
\partial \bar{W}_0 / \partial r_0 + \Sigma ( \partial \bar{W}_j / \partial r_j
) = 0
\end{eqnarray}
further, if the last term is equal to zero, we have
\begin{eqnarray}
\partial \bar{W}_0 / \partial t = 0.
\end{eqnarray}
where, when the time $t$ is only the independent variable, the
$\partial \bar{W}_0 / \partial t$ will become the $d \bar{W}_0 / d
t$ .

The above means that the power-like density $N_0$ will be variable
in the case for coexistence of gravitational field and
electromagnetic field, although the $\bar{N}_0$ is the scalar
invariant under the octonion transformation. And the potential,
strength, velocity curl, or torque density, etc. have the influence
on the energy-like continuity equation in the gravitational field
and electromagnetic field.

\section{CONCLUSIONS}

In the quaternion compounding space, the inferences about
conservation laws depend on the combinations of physical
definitions. By means of definition combination of radius vector and
velocity, the gravitational potential is found to have the influence
on the speed of light in the gravitational field. With the
definition combination of the velocity and linear momentum, it is
found that the mass density is variable, and the gravitational
strength has the impact on the conservation of mass. Depending on
the definition combination of the angular momentum and velocity, we
obtain that the spin density and energy density both are variable,
and the gravitational strength has the impact on the conservation of
energy.

In the octonion compounding space, the results about the mechanics
invariants of electromagnetic field depend on the definition
combinations, in the case for coexistence of the gravitational field
and electromagnetic field. With the definitions of linear momentum
and velocity, the mass density will be variable and the conservation
of mass will be changed with the strength and velocity curl. From
the definitions of angular momentum and velocity, the spin density
and the energy density will be variable for the influence of the
potential and velocity curl etc., while the conservation of spin and
the conservation of energy will be changed with the impact of the
velocity curl and strength etc.

In the octonion compounding space, the results about the scalar
invariants with the velocity curl depend on the definition
combinations of physical quantities, in the case for coexistence of
gravitational field and electromagnetic field. The mechanical
invariants are different from the electric invariants. The former is
related to the mass, and the latter is associated with the charge.
With the definitions of linear momentum and velocity, the charge
density will be variable and the conservation of charge will be
changed with the strength and velocity curl. From the definitions of
angular momentum and velocity, the spin magnetic moment density and
the energy-like density both will be variable for the influence of
the potential and velocity curl etc., while the conservation of spin
magnetic moment and the conservation of energy-like will be changed
with the impact of the velocity curl and strength etc. in the
gravitational field and electromagnetic field.

It should be noted that the study for scalar invariants of the
electromagnetic field and gravitational field examined only one
simple case with very weak strength and low velocity curl in the
gravitational field and electromagnetic field. Despite its
preliminary character, this study can clearly indicate the strength
and velocity curl in the gravitational field and electromagnetic
field have the limited influence on the scalar invariants. For the
future studies, the related investigation will concentrate on only
the predictions of scalar invariants in the strong strength with
high velocity curl in the gravitational field and electromagnetic
field.

\begin{acknowledgments}
This project was supported partially by the National Natural Science
Foundation of China under grant number 60677039, Science \&
Technology Department of Fujian Province of China under grant number
2005HZ1020 and 2006H0092, and Xiamen Science \& Technology Bureau of
China under grant number 3502Z20055011.
\end{acknowledgments}

\end{document}